\documentclass[twocolumn,floatfix,prb,superscriptaddress]{revtex4}
\usepackage{graphicx}
\usepackage{amsmath}
\usepackage{amssymb}
\usepackage{bm}

\begin{document}

\title{Time-delay matrix, midgap spectral peak, and thermopower of an Andreev billiard}
\author{M. Marciani}
\affiliation{Instituut-Lorentz, Universiteit Leiden, P.O. Box 9506, 2300 RA Leiden, The Netherlands}
\author{P. W. Brouwer}
\affiliation{Dahlem Center for Complex Quantum Systems and Fachbereich Physik, Freie Universit\"{a}t Berlin, Arnimallee 14, 14195 Berlin, Germany}
\author{C. W. J. Beenakker}
\affiliation{Instituut-Lorentz, Universiteit Leiden, P.O. Box 9506, 2300 RA Leiden, The Netherlands}
\date{May 2014}
\begin{abstract}
We derive the statistics of the time-delay matrix (energy derivative of the scattering matrix) in an ensemble of superconducting quantum dots with chaotic scattering (Andreev billiards), coupled ballistically to $M$ conducting modes (electron-hole modes in a normal metal or Majorana edge modes in a superconductor). As a first application we calculate the density of states $\rho_0$ at the Fermi level. The ensemble average $\langle\rho_0\rangle=\delta_0^{-1}M[\max(0,M+2\alpha/\beta)]^{-1}$ deviates from the bulk value $1/\delta_0$ by an amount depending on the Altland-Zirnbauer symmetry indices $\alpha,\beta$. The divergent average for $M=1,2$ in symmetry class D ($\alpha=-1$, $\beta=1$) originates from the mid-gap spectral peak of a closed quantum dot, but now no longer depends on the presence or absence of a Majorana zero-mode. As a second application we calculate the probability distribution of the thermopower, contrasting the difference for paired and unpaired Majorana edge modes. 
\end{abstract}
\maketitle

\section{Introduction}
\label{intro}

\begin{figure}[tb]
\centerline{\includegraphics[width=0.8\linewidth]{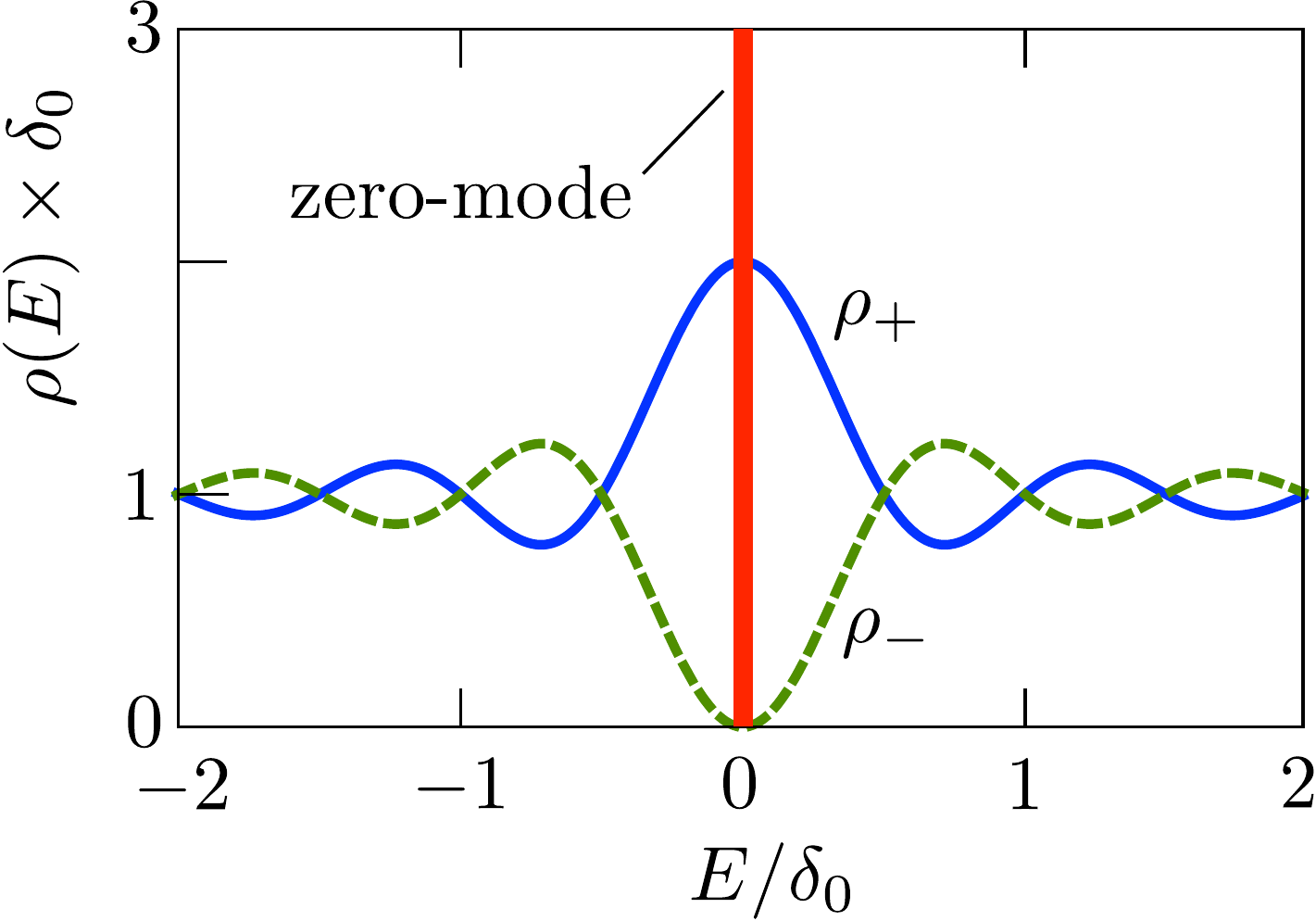}}
\caption{Ensemble-averaged density of states \eqref{rhopmresult} of an Andreev billiard in symmetry class C ($\rho_-$, dashed curve), or class D without a Majorana zero-mode ($\rho_+$, solid curve). The class D billiard with a Majorana zero-mode has the smooth density of states $\rho_-$ together with the delta-function contribution from the zero-mode. In this paper we investigate how the midgap spectral peak or dip evolves when the billiard is opened via a ballistic point contact to a metallic reservoir. We find that the distinction between class C and D remains, but the signature of the Majorana zero-mode is lost.
}
\label{fig_spectralpeak}
\end{figure}

A semiconductor quantum dot feels the proximity to a superconductor even when a magnetic field has closed the excitation gap that would open in zero magnetic field: The average density of states has either a peak or a dip,\cite{Alt97}
\begin{equation}
\rho_{\pm}( E)=\delta_{0}^{-1}\pm\frac{\sin(2\pi E/\delta_0)}{2\pi E},\label{rhopmresult}
\end{equation}
see Fig.\ \ref{fig_spectralpeak}, within a mean level spacing $\delta_0$ from the Fermi level at $ E=0$ (in the middle of the superconducting gap). The appearance of a midgap spectral peak or dip distinguishes the two symmetry classes C (dip, when spin-rotation symmetry is preserved) and D (peak, spin-rotation symmetry is broken by strong spin-orbit coupling). These Altland-Zirnbauer symmetry classes exist because of the $\pm E$ electron-hole symmetry in a superconductor, and are a late addition to the Wigner-Dyson symmetry classes conceived in the 1960's to describe universal properties of nonsuperconducting systems.\cite{handbook}

Electron-hole symmetry in the absence of spin-rotation symmetry allows for a nondegenerate level at $ E=0$, a socalled Majorana zero-mode.\cite{Vol99,Rea00} The class-D spectral peak is then converted into a dip, $\rho_{+}\rightarrow\rho_{-}+\delta( E)$, such that the integrated density of states remains the same as without the zero-mode.\cite{Boc00,Iva02} The entire spectral weight of this Fermi-level anomaly is $1/2$, consistent with the notion that a Majorana zero-mode is a half-fermion.\cite{Jac76}

Here we study what happens if the quantum dot is coupled to $M$ conducting modes, so that the discrete spectrum of the closed system is broadened into a continuum. We focus on the strong-coupling limit, typically realized by a ballistic point contact, complementing earlier work on the limit of weak coupling by a tunnel barrier or a localized conductor.\cite{Bag12,Liu12,Nev13,Skv13,Sta13,Sau13,Ios13,Iva13} The simplicity of the strong-coupling limit allows for an analytical calculation using random-matrix theory of the entire probability distribution of the Fermi-level density of states --- not just the ensemble average. Using the same random-matrix approach we also calculate the probability distribution of the thermopower of the quantum dot, which is nonzero in spite of electron-hole symmetry when the superconductor contains gapless Majorana edge modes.\cite{Hou13}

The key technical ingredient that makes these calculations possible is the joint probability distribution of the scattering matrix $S$ and the time-delay matrix $Q=-i\hbar S^{\dagger}dS/d E$, in the limit $ E\rightarrow 0$. This is known for the Wigner-Dyson ensembles,\cite{Bro97} and here we extend that to the Altland-Zirnbauer ensembles. The Fermi-level density of states then follows directly from the trace of $Q$, while the thermopower requires also knowledge of the statistics of $S$. We find that these probability distributions depend on the symmetry class (C or D), and on the number $M$ of conducting modes, but are the same irrespective of whether the quantum dot contains a Majorana zero-mode or not. A previous calculation\cite{Nev13} had found that the density-of-states signature of a Majorana zero-mode becomes less evident when the quantum dot is coupled by a tunnel barrier to the continuum. We conclude that ballistic coupling completely removes any trace of the Majorana zero-mode in the density of states, as well as in the thermopower --- but not, we hasten to add, in the Andreev conductance.\cite{Bee11}

The outline of the paper is as follows. In the next section we present the geometry of an ``Andreev billiard'',\cite{Bee05} a semiconductor quantum dot with Andreev reflection from a superconductor and a point-contact coupling to a metallic conductor. (Systems of this type have been studied experimentally, for example in Refs.\ \onlinecite{Dir11,Lee12,Cha13}.) We derive a formula relating the thermopower to the scattering matrix $S$ and time-delay matrix $Q$, in a form which is suitable for a random-matrix approach. The distribution of the transmission eigenvalues $T_n$ of $S$ was already derived in Ref.\ \onlinecite{Dah10}; what we need additionally is the distribution of the eigenvalues $D_n$ of $Q$ (the delay times), which we present in Sec.\ \ref{circular}. The distributions of the Fermi-level density of states and thermopower are given in Secs.\ \ref{rho0distribution} and \ref{thermodist}, respectively. We conclude in Sec.\ \ref{conclude}.

\section{Scattering formula for the thermopower}
\label{thermop}

\begin{figure}[tb]
\centerline{\includegraphics[width=0.9\linewidth]{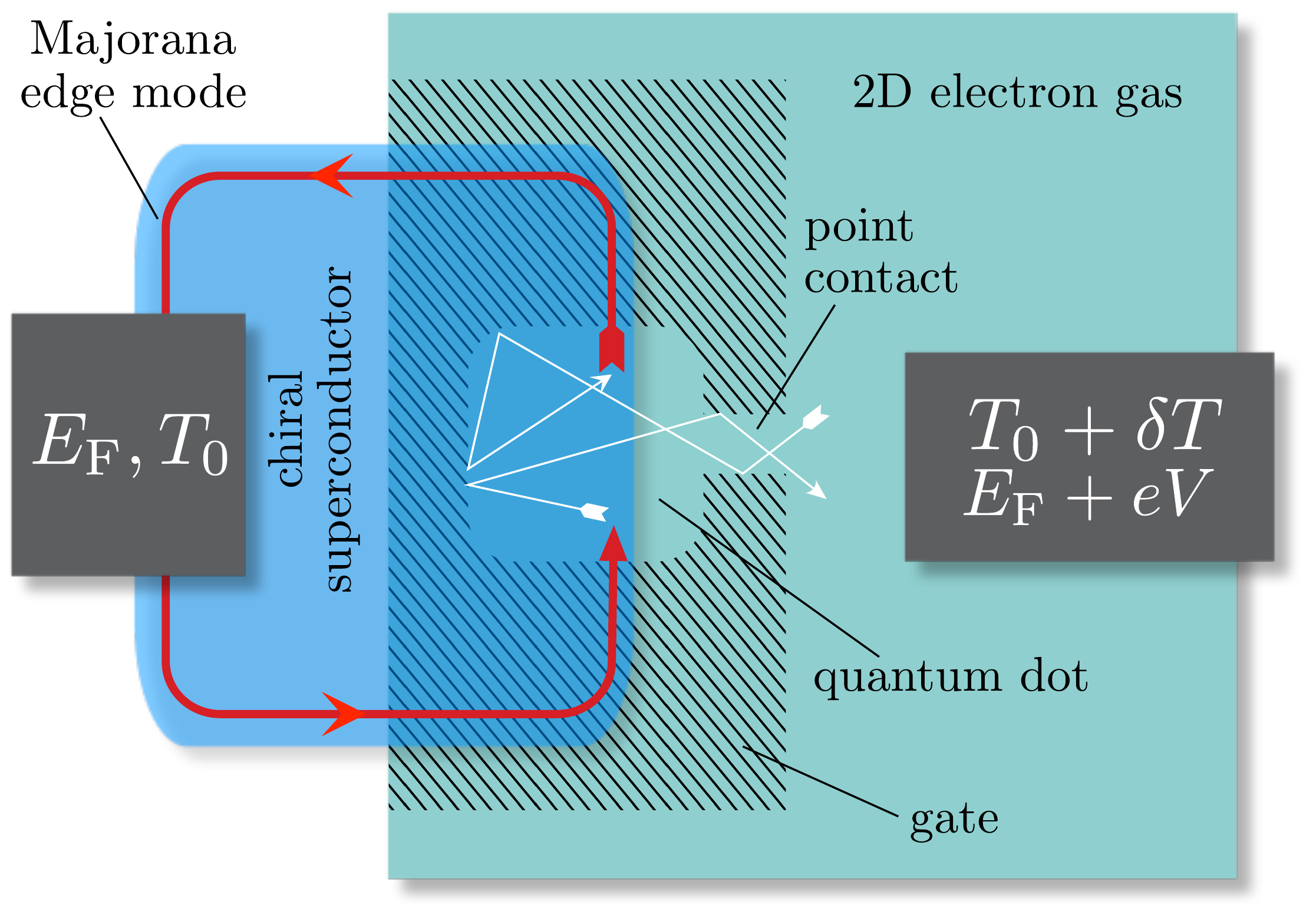}}
\caption{Andreev-billiard geometry to measure the thermopower ${\cal S}$ of a semiconductor quantum dot coupled to chiral Majorana modes at the edge of a topological superconductor. A temperature difference $\delta T$ induces a voltage difference $V=-{\cal S}\delta T$ under the condition that no electrical current flows between the contacts. For a random-matrix theory we assume that the Majorana modes are uniformly mixed with the modes in the point contact, by chaotic scattering events in the quantum dot.
}
\label{fig_chiral}
\end{figure}

We study the thermopower of a quantum dot connecting a two-dimensional topological superconductor and a semiconductor two-dimensional electron gas (see Fig.\ \ref{fig_chiral}). In equilibrium the normal-metal contact and the superconducting contact have a common temperature $T_0$ and chemical potential $E_{\rm F}$. Application of a temperature difference $\delta T$ induces a voltage difference $V$ at zero electrical current. The ratio ${\cal S}=-V/\delta T$ is the thermopower or Seebeck coefficient. 

In the low-temperature limit $\delta T\ll T_0\rightarrow 0$ the thermopower is given by the Cutler-Mott formula,\cite{Cut69}
\begin{equation}
{\cal S}/{\cal S}_{0}=-\lim_{E\rightarrow 0}\frac{1}{G}\frac{dG}{dE},\;\;{\cal S}_{0}=\frac{\pi^{2}k_{\rm B}^{2}T_{0}}{3e},\label{CutlerMott}
\end{equation}
in terms of the electrical conductance $G(E)$ near the Fermi level ($E=0$). See Ref.\ \onlinecite{Hou13} for a demonstration that this relationship, originally derived for normal metals, still holds when one of the contacts is superconducting and $G$ is the Andreev conductance.

Without gapless Majorana modes in the superconductor the Andreev conductance is an even function of $E$, so the ratio ${\cal S}/{\cal S}_0$ vanishes in the low-temperature limit. For that reason, with some exceptions,\cite{Kal12,Oza14} most studies of the effect of a superconductor on thermo-electric transport take a three-terminal geometry, where the temperature difference is applied between two normal contacts and the conductance is not so constrained.\cite{Cla96,Eom98,Sev00,Dik02,Par03,Vol05,Vir04,Sri05,Jac10,Mac13} As pointed out by Hou, Shtengel, and Refael,\cite{Hou13} Majorana edge modes break the $\pm E$ symmetry of the conductance allowing for thermo-electricity in a two-terminal geometry --- even if they themselves carry only heat and no charge.

In a random-matrix formulation of the problem two matrices enter, the scattering matrix at the Fermi level $S_0\equiv S(E=0)$ and the Wigner-Smith time-delay matrix\cite{Wig55,Smi60,Fyo10}
\begin{equation}
Q=-i\hbar \lim_{E\rightarrow 0}S^{\dagger}\frac{dS}{dE}.\label{Qdef} 
\end{equation}
Before proceeding to the random-matrix theory, we first express the thermopower in terms of these two matrices. The existing expressions in the literature \cite{Lan98,God99} cannot be directly applied for this purpose, since they do not incorporate Andreev reflection processes.

The Andreev conductance is given by \cite{Tak92}
\begin{equation}
G(E)/G_{0}=\tfrac{1}{2}N-{\rm Tr}\,r_{ee}^{\vphantom{\dagger}}(E)r_{ee}^{\dagger}(E)+{\rm Tr}\,r_{he}^{\vphantom{\dagger}}(E)r_{he}^{\dagger}(E),\label{Gformula}
\end{equation}
in terms of the matrix of reflection amplitudes 
\begin{equation}
r=\begin{pmatrix}
r_{ee}&r_{eh}\\
r_{he}&r_{hh}
\end{pmatrix}\label{rdef}
\end{equation}
for electrons and holes injected via a point contact into the quantum dot. The submatrix $r_{ee}$ describes normal reflection (from electron back to electron), while $r_{he}$ describes Andreev reflection (from electron to hole, induced by the proximity effect of the superconductor that interfaces with the quantum dot). The conductance quantum is $G_{0}=e^2/h$ and $N$ is the total number of modes in the point contact (counting spin and electron-hole degrees of freedom), so $r$ has dimension $N\times N$. 

Without edge modes in the superconductor, the reflection matrix $r$ would be unitary at energies $E$ below the superconducting gap. In that case one can simplify Eq.\ \eqref{Gformula} as $G/G_0=2\,{\rm Tr}\,r_{he}^{\vphantom{\dagger}}r_{he}^{\dagger}$. Because of the gapless edge modes the more general formula \eqref{Gformula} is needed, which does not assume unitarity of $r$. 

Equivalently, Eq.\ \eqref{Gformula} may be written in terms of the full unitary scattering matrix $S(E)$,
\begin{equation}
G(E)/G_{0}=\tfrac{1}{2}N-\tfrac{1}{2}{\rm Tr}\,{\cal P}\tau_{z}S(E){\cal P}(1+\tau_{z})S^{\dagger}(E),\label{GeSSrelation}
\end{equation}
where the Pauli matrix $\tau_z$ acts on the electron-hole degree of freedom and ${\cal P}$ projects onto the modes at the point contact:
\begin{align}
&S=\begin{pmatrix}
r&t'\\
t&r'
\end{pmatrix},\;\;{\cal P}\tau_{z}=\begin{pmatrix}
\tau_{z}&0\\
0&0
\end{pmatrix}.\label{projections}
\end{align}
The off-diagonal matrix blocks $t,t'$ couple the $N'$ Majorana edge modes to the $N$ electron-hole modes in the point contact, mediated by the quasibound states in the quantum dot. The incoming and outgoing Majorana edge modes are coupled by the $N'\times N'$ submatrix $r'$.

Electron-hole symmetry in class D is most easily accounted for by first making a unitary transformation from $S$ to
\begin{equation}
S'= \begin{pmatrix}
 U&0\\
0& U
\end{pmatrix} S\begin{pmatrix}
 U^{\dagger}&0\\
0& U^{\dagger}
\end{pmatrix},\;\; U=\sqrt{\tfrac{1}{2}}\begin{pmatrix}
1&1\\
i&-i
\end{pmatrix}.\label{SMajoranatransf}
\end{equation}
In this socalled Majorana basis \cite{note1} the electron-hole symmetry relation reads
\begin{equation}
S'(E)=S'^{\ast}(-E).\label{SehsymM}
\end{equation}
The Pauli matrix $\tau_z$ transforms into $\tau_y$, so the conductance is given in the Majorana basis by
\begin{equation}
G(E)/G_{0}=\tfrac{1}{2}N-\tfrac{1}{2}{\rm Tr}\,{\cal P}\tau_{y}S'(E){\cal P}(1+\tau_{y})S'^{\dagger}(E).\label{GeSSMrelation}
\end{equation}
In what follows we will omit the prime, for ease of notation.

To first order in $E$ the energy dependence of the scattering matrix is given by
\begin{equation}
S(E)=S_{0}[1+iE\hbar^{-1} Q+{\cal O}(E^2)].\label{SQdef}
\end{equation}
Unitarity and electron-hole symmetry together require that $S_0$ is real orthogonal and $Q$ is real symmetric, both in the Majorana basis. The conductance, still to first order in $E$, then takes the form
\begin{align}
&G(E)/G_{0}=\tfrac{1}{2}N-\tfrac{1}{2}{\rm Tr}\,{\cal P}\tau_{y}S_0{\cal P}(1+\tau_{y})S_{0}^{\rm T}\nonumber\\
&\;-\tfrac{1}{2}iE\hbar^{-1}\,{\rm Tr}\,{\cal P}\tau_{y}S_{0}\bigl[Q{\cal P}(1+\tau_{y})-{\cal P}(1+\tau_{y})Q\bigr]S_{0}^{\rm T}.\label{GlinearinE}
\end{align}

Since ${\rm Tr}\,{\cal P}\tau_{y}X$ vanishes for any symmetric matrix $X$, we can immediately set some of the traces in Eq.\ \eqref{GlinearinE} to zero:
\begin{align}
G(E)/G_{0}={}&\tfrac{1}{2}N-\tfrac{1}{2}{\rm Tr}\,{\cal P}\tau_{y}S_0{\cal P}\tau_{y}S_{0}^{\rm T}\nonumber\\
&-\tfrac{1}{2}iE\hbar^{-1}\,{\rm Tr}\,{\cal P}\tau_{y}S_{0}\bigl(Q{\cal P}-{\cal P}Q\bigr)S_{0}^{\rm T}.\label{GlinearinEsimpler}
\end{align}
The resulting thermopower is
\begin{equation}
{\cal S}/{\cal S}_{0}=i\hbar^{-1}\frac{{\rm Tr}\,{\cal P}\tau_{y}S_{0}(Q{\cal P}-{\cal P}Q)S_{0}^{\rm T}}{N-{\rm Tr}\,{\cal P}\tau_{y}S_0{\cal P}\tau_{y}S_{0}^{\rm T}},\label{Presult}
\end{equation}
in the Majorana basis. Equivalently, in the electron-hole basis one has
\begin{equation}
{\cal S}/{\cal S}_{0}=i\hbar^{-1}\frac{{\rm Tr}\,{\cal P}\tau_{z}S_{0}(Q{\cal P}-{\cal P}Q)S_{0}^{\dagger}}{N-{\rm Tr}\,{\cal P}\tau_{z}S_0{\cal P}\tau_{z}S_{0}^{\dagger}}.\label{Presulteh}
\end{equation}

This scattering formula for the thermopower is a convenient starting point for a random-matrix calculation. Notice that the commutator of $Q$ and ${\cal P}$ in the numerator ensures a vanishing thermopower in the absence of gapless modes in the superconductor, because then the projector ${\cal P}$ is just the identity.

\section{Delay-time distribution in the Altland-Zirnbauer ensembles}
\label{circular}

\begin{table}
\centering
\begin{tabular}{ | l || c | c | }
\hline
symmetry class &  C & D \\ \hline
pair potential & spin-singlet d-wave & spin-triplet p-wave \\ \hline
canonical basis & electron-hole & Majorana \\ \hline
$S$-matrix elements\ & quaternion &  real \\ \hline
$S$-matrix space & symplectic & orthogonal \\ \hline
circular ensemble & CQE & CRE \\ \hline
\qquad\qquad $d_T$ & $4$ &  $1$ \\ \hline
\qquad\qquad $d_E$ & $2$ &  $1$ \\ \hline
\qquad\qquad $\alpha$ & $2$&$-1$\\ \hline
\qquad\qquad $\beta$ &$4$&$1$\\ \hline
\end{tabular}
\caption{The two Altland-Zirnbauer symmetry classes that support chiral Majorana edge modes, with d-wave pairing (class C) or p-wave pairing (class D). The ``canonical basis'' is the basis in which the scattering matrix elements are quaternion (class C) or real (class D). The degeneracies $d_T$ and $d_E$ refer to transmission eigenvalues and energy eigenvalues, respectively.  The $\alpha$ and $\beta$ parameters determine the exponents in the probability distributions \eqref{PTn} and \eqref{Pgamman} of the transmission eigenvalues and inverse delay times.
}
\label{table_CD}
\end{table}

Chaotic scattering in the quantum dot mixes the $N'$ Majorana edge modes with the $N$ electron-hole modes in the point contact. The assumption that the mixing uniformly covers the whole available phase space produces one of the circular ensembles of random-matrix theory, distinguished by fundamental symmetries that restrict the available phase space.\cite{Zir11} Two Altland-Zirnbauer symmetry classes support chiral Majorana modes at the edge of a two-dimensional superconductor,\cite{Ryu10,Has10,Qi11} corresponding to spin-singlet d-wave pairing (symmetry class C) or spin-triplet p-wave pairing (symmetry class D). Time-reversal symmetry is broken in both, in class C there is electron-hole symmetry as well as spin-rotation symmetry, while in class D only electron-hole symmetry remains. (See Table \ref{table_CD}.)

The uniformity of the circular ensembles is expressed by the invariance
\begin{equation}
P[S(E)]=P[U\cdot S(E)\cdot U']\label{PSEinvariance}
\end{equation}
of the distribution functional $P[S(E)]$ upon multiplication of the scattering matrix by a pair of energy-independent matrices $U,U'$, restricted by symmetry to a subset of the full unitary group: In class C they are quaternion symplectic\cite{note4} in the electron-hole basis (circular quaternion ensemble, CQE), while in class D they are real orthogonal in the Majorana basis (circular real ensemble, CRE).

The unitary invariance \eqref{PSEinvariance} of the Wigner-Dyson scattering matrix ensembles was postulated in Ref.\ \onlinecite{Wig51} and derived from the corresponding Hamiltonian ensembles in Ref.\ \onlinecite{Bro99}. We extend the derivation to the Altland-Zirnbauer ensembles in App.\ \ref{invariance_proof}. The key step in this extension is to ascertain that the class-D unitary invariance applies to $U,U'$ in the full orthogonal group --- without any restriction on the sign of the determinant. 

For the thermopower statistics we need the joint distribution $P(S_0,Q)$ of Fermi-level scattering matrix and time-delay matrix. The invariance \eqref{PSEinvariance} implies $P(S_0,Q)=P(-1,Q)$ (take $U=-S_0^\dagger$, $U'=1$), so $Q$ is statistically independent of $S_0$ and the two matrices can be considered separately.\cite{Bro97,note2} 

The uniform distribution of $S_0$ in the symplectic group (CQE, class C) or orthogonal group (CRE, class D) directly gives the probability distribution of the transmission eigenvalues $T_{n}\in[0,1]$ of quasiparticles from the normal metal into the superconductor. [These are the quantities that determine the thermal conductance $\propto\sum_{n}T_{n}$, not the electrical conductance \eqref{Gformula}.] For a transmission matrix of dimension $N'\times N$ there are $N_{\rm min}={\rm min}\,(N,N')$ nonzero transmission eigenvalues, fourfold degenerate ($d_{T}=4$) in class C and nondegenerate ($d_{T}=1$) in class D. The $N_{\rm min}/d_{T}$ distinct $T_{n}$'s have probability distribution\cite{Dah10}
\begin{align}
P(\{T_{n}\})\propto{}&\prod_{k}T_{k}^{\beta|\delta N|/2}T_{k}^{-1+\beta/2}(1-T_{k})^{\alpha/2}\nonumber\\
&\times\prod_{i<j}|T_{i}-T_{j}|^{\beta},\label{PTn}
\end{align}
with $\delta N=(N-N')/d_{T}$ and parameters $\alpha,\beta$ listed in Table \ref{table_CD}.\cite{Bro00}

The Hermitian positive-definite matrix $Q$ has dimension ${\cal M}\times{\cal M}$ with ${\cal M}=N+N'$. Its eigenvalues $D_n>0$ are the delay times, and $\gamma_n\equiv 1/D_n$ are the corresponding rates. The degeneracy $d_T$ of the $D_n$'s is the same as that of the $T_n$'s. The derivation of the distribution $P(\gamma_1,\gamma_2,\ldots\gamma_M)$ of the $M={\cal M}/d_T$ distinct delay rates is given in App.\ \ref{detailsdelaytime}, for all four Altland-Zirnbauer symmetry classes: C, D without time-reversal symmetry and CI, DIII with time-reversal symmetry. The result is
\begin{align}
P(\{\gamma_{n}\})\propto{}&\prod_{k}\Theta(\gamma_k)\gamma_{k}^{\alpha+M\beta/2}\exp\left(-\tfrac{1}{2}\beta t_{0}\gamma_{k}\right)\nonumber\\
&\times\prod_{i<j}|\gamma_{i}-\gamma_{j}|^{\beta}.\label{Pgamman}
\end{align}
The unit step function $\Theta(\gamma)$ ensures that the probability vanishes if any $\gamma_{n}$ is negative. The characteristic time $t_0$ is defined by
\begin{equation}
t_0=\frac{d_{E}}{d_{T}}\frac{2\pi\hbar}{\delta_0},\label{t0def}
\end{equation}
in terms of the average spacing $\delta_0$ of $d_{E}$-fold degenerate energy levels in the isolated quantum dot.\cite{note6} For $\alpha=0$ and $d_{E}=d_{T}$ we recover the result of Ref.\ \onlinecite{Bro97} for the Wigner-Dyson ensembles.

The difference between the Altland-Zirnbauer and Wigner-Dyson ensembles manifests itself in a nonzero value of $\alpha$ and in a difference in the degeneracies $d_E$ and $d_T$ of energy and transmission eigenvalues (see Table \ref{table_CD}). One has $d_T=d_E$ in the absence of particle-hole symmetry or when the particle-hole conjugation operator ${\cal C}$ squares to $+1$; when ${\cal C}^2=-1$ one has $d_T=2d_E$.\cite{note8}

Already at this stage we can conclude that the thermopower distribution in the circular ensemble does not depend on the presence or absence of Majorana zero-modes inside the quantum dot, for example, bound to the vortex core in a chiral p-wave superconductor.\cite{Vol99,Rea00} The parity of the number $n_{\rm M}$ of Majorana zero-modes fixes the sign of the determinant of the orthogonal class-D scattering matrix,
\begin{equation}
{\rm Det}\,S_0=(-1)^{n_{\rm M}}.\label{DetSonM} 
\end{equation}
The unitary invariance \eqref{PSEinvariance} of the CRE implies, on the one hand, that $P(S_0,Q)$ is unchanged under the transformation $S_0\mapsto US_0$, $U={\rm diag}\,(-1,1,1,\ldots 1)$, that inverts the sign of ${\rm Det}\,S_0$. (Here we make essential use of the fact that Eq.\ \eqref{PSEinvariance} in class D applies to the full orthogonal group.) On the other hand, the same transformation leaves the thermopower \eqref{Presult} unaffected, provided we assign the first matrix element to a superconducting edge mode (so ${\cal P}\tau_y$ commutes with $U$).

\section{Fermi-level anomaly in the density of states}
\label{rho0distribution}

\subsection{Analytical calculation}
\label{rho0analytical}

A striking difference between the Wigner-Dyson and Altland-Zirnbauer ensembles appears when one considers the density of states at the Fermi level $\rho_0$, related to the time-delay matrix by
\begin{equation}
\rho_0=\frac{1}{2\pi\hbar}\frac{d_{T}}{d_{E}}\,\sum_{n=1}^{M}D_n.\label{rho0def}
\end{equation}
(The factor $d_T/d_E$ is needed because delay times and energy levels may have a different degeneracy. The density of states counts degenerate levels once.) In the Wigner-Dyson ensembles the average density of states equals exactly $1/\delta_{0}$, independent of the symmetry index $\beta$ and of the number of channels $M$ that couple the discrete spectrum inside the quantum dot to the continuum outside.\cite{Bro97,Lyu77} 

In the Altland-Zirnbauer ensembles, instead, we find from Eq.\ \eqref{Pgamman} that\cite{note7}
\begin{align}
&\delta_0\langle\rho_0\rangle =\frac{1}{t_{0}}\left\langle\sum_{n=1}^{M}D_n\right\rangle=\frac{M}{\max(0,M+2\alpha/\beta)}\nonumber\\
&=\begin{cases}
M/(M+1)& \text{in class C for any}\;\;M\geq 1,\\
M/(M-2)& \text{in class D for}\;\;M\geq 3,\\
\infty& \text{in class D for}\;\;M=1,2.
\end{cases}\label{rho0result}
\end{align}
It is known\cite{Alt97,Iva02,Bag12,Liu12,Nev13,Skv13,Sta13,Sau13,Ios13,Iva13} that the tunneling density of states of a superconducting quantum dot with broken time-reversal symmetry, weakly coupled to the outside, has a Fermi-level anomaly consisting of a narrow dip in symmetry class C and a narrow peak in class D. Eq.\ \eqref{rho0result} shows the effect of level broadening upon coupling via $M$ channels to the continuum. For $M\rightarrow\infty$ the normal-state result $1/\delta_{0}$ is recovered, but for small $M$ the Fermi-level anomaly persists.

\begin{figure}[tb]
\centerline{\includegraphics[width=0.9\linewidth]{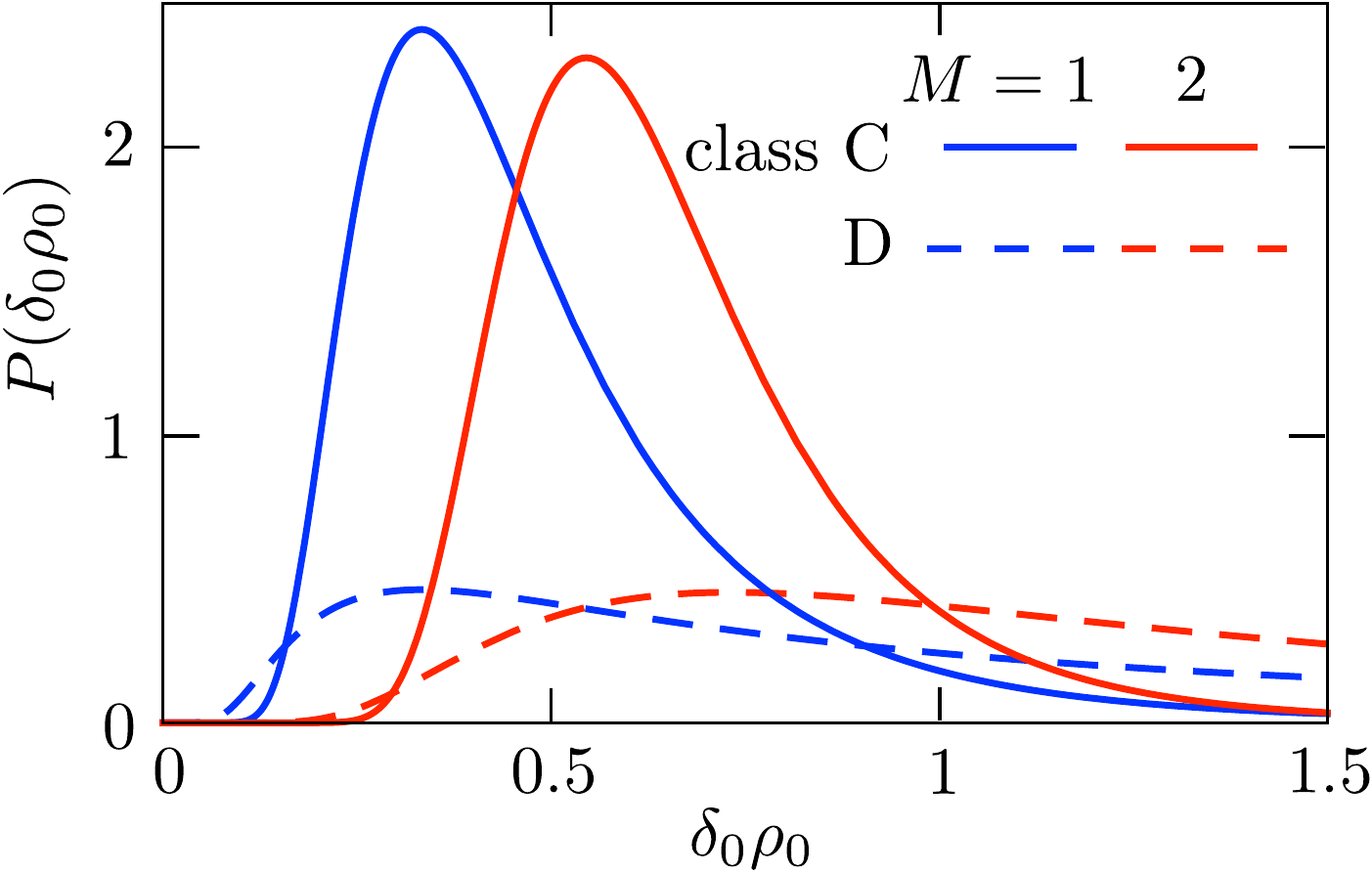}}
\caption{Probability distributions of the Fermi-level density of states, for $M=1$ and $M=2$ modes coupling the quantum dot to the continuum, in symmetry classes C and D. The ensemble average diverges for class D, see Eq.\ \eqref{rho0result}.
}
\label{rho0plot}
\end{figure}

For $M=1,2$ the average density of states in class D diverges, because of a long tail in the probability distribution of $\kappa\equiv\delta_0\rho_0$:
\begin{equation}
P_{\rm D}(\kappa)=\begin{cases}
(2\pi)^{-1/2}\kappa^{-3/2}e^{-(2\kappa)^{-1}}&{\rm for}\;\;M=1,\\
\kappa^{-3}(2+\kappa)e^{-2/\kappa}&{\rm for}\;\;M=2.
\end{cases}\label{PDrho0}
\end{equation}
See Fig.\ \ref{rho0plot} for a plot and a comparison with the class-C distribution, that has a finite average for alle $M$.

The result \eqref{PDrho0} holds irrespective of the sign of ${\rm Det}\,S_0$, in other words, the statistics of the Fermi-level anomaly in the CRE does not depend on the presence or absence of an unpaired Majorana zero-mode in the quantum dot. As we remarked at the end of the previous section, in connection with the thermopower, this is a direct consequence of the unitary invariance \eqref{PSEinvariance} of the circular ensemble.

\subsection{Numerical check}
\label{rho0numerical}

As check on our analytical result we have calculated $P(\rho_0)$ numerically from the Gaussian ensemble of random Hamiltonians. We focus on symmetry class D, where we can test in particular for the effect of a Majorana zero-mode.

The Hamiltonian $H$ is related to the scattering matrix $S( E)$ by the Weidenm\"{u}ller formula,\cite{Guh98,Bee97}
\begin{align}
&S( E)=\frac{1+i\pi W^{\dagger}(H- E)^{-1}W}{1-i\pi W^{\dagger}(H- E)^{-1}W}\nonumber\\
&\quad=1+2\pi iW^{\dagger}(H_{\rm eff}- E)^{-1}W,\;\;H_{\rm eff}=H-i\pi WW^{\dagger}.\label{SHeq}
\end{align}
The $M_0\times M$ matrix $W$ couples the $M_0$ energy levels in the quantum dot to $M\ll M_0$ scattering channels. Ballistic coupling corresponds to
\begin{equation}
W_{nm}=\delta_{nm}\sqrt{M_{0}\delta_0}/\pi.\label{Wnmdef}
\end{equation}

The density of states is determined by the scattering matrix via\cite{Akk91}
\begin{equation}
\rho( E)=-\frac{i}{2\pi}\frac{d}{d E}\ln{\rm Det}\,S( E).\label{rhovarepsilonlogDetS}
\end{equation}
From Eqs.\ \eqref{SHeq} and \eqref{rhovarepsilonlogDetS} we obtain an expression for the Fermi-level density of states in terms of the Hamiltonian,
\begin{equation}
\rho_0={\rm Tr}\,\biggl(\bigl[1-2\pi iW^{\dagger}(H_{\rm eff}^{\dagger})^{-1}W\bigr]W^{\dagger}H_{\rm eff}^{-2}W\biggr).\label{QdefH}
\end{equation}

In the Majorana basis the class-D Hamiltonian is purely imaginary, $H=iA$, with $A$ a real antisymmetric matrix. The Gaussian ensemble has probability distribution\cite{Iva02,Mehta}
\begin{equation}
P(A)\propto\prod_{n>m}\exp\left(-\frac{\pi^{2}A_{nm}^{2}}{2M_0\delta_0^{2}}\right).\label{GaussEns}
\end{equation}
The dimensionality of $A$ is odd if the quantum dot contains an unpaired Majorana zero-mode, otherwise it is even.

\begin{figure}[tb]
\centerline{\includegraphics[width=1\linewidth]{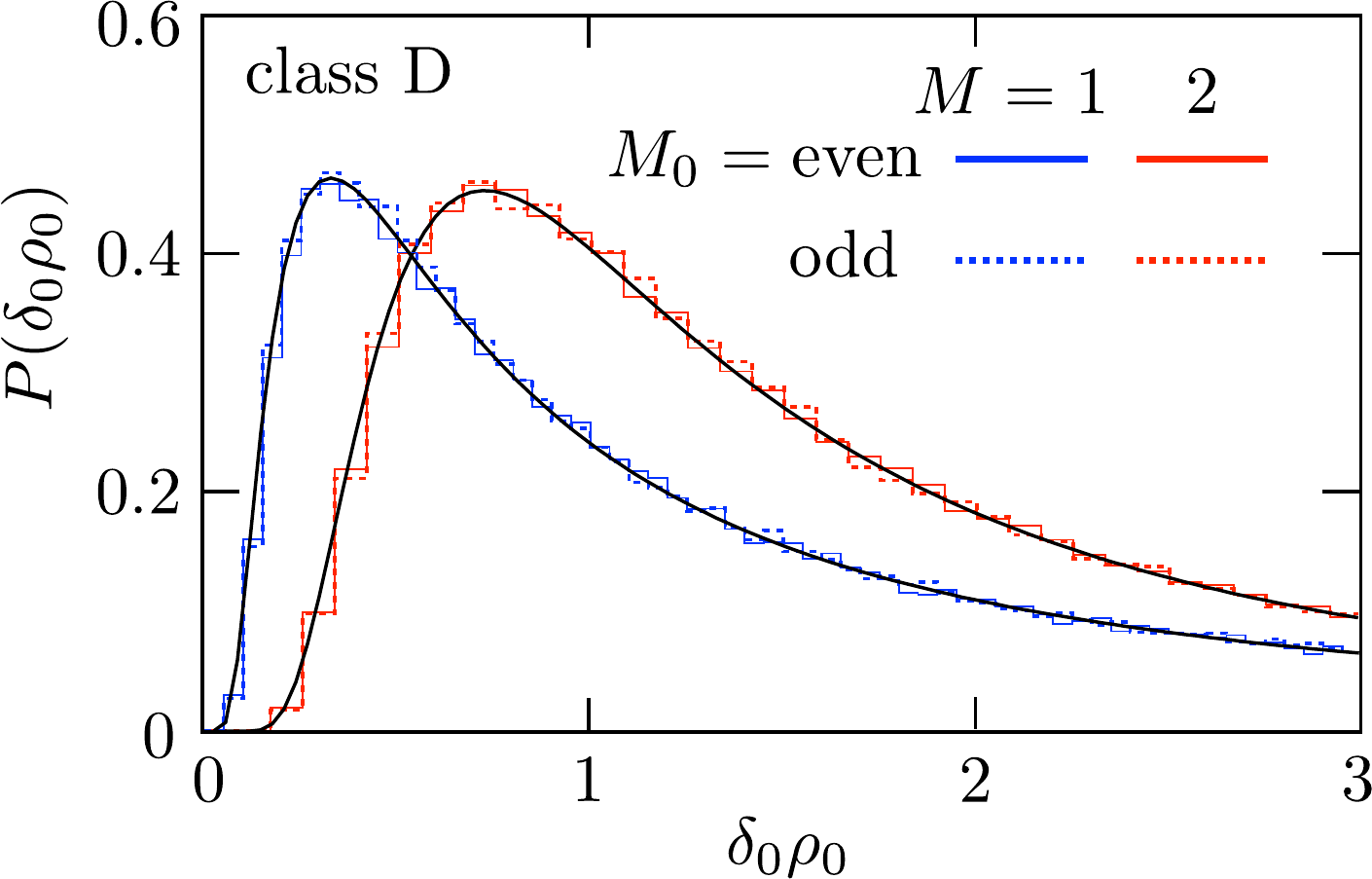}}
\caption{Histograms: Probability distributions of the Fermi-level density of states in symmetry class D for $M=1$, $M_0=140,141$ and $M=2$, $M_0=200,201$, calculated numerically from Eq.\ \eqref{QdefH} by averaging the Hamiltonian over the Gaussian ensemble. For each dimensionality $M$ of the scattering matrix we compare an even-dimensional Hamiltonian, without a Majorana zero-mode, to an odd-dimensional Hamiltonian with a zero-mode. The black curve is the analytical result \eqref{PDrho0} for the circular scattering matrix ensemble, predicting no effect from the Majorana zero-mode for this case of ballistic coupling. Notice that there is no fit parameter in this comparison between numerics and analytics.
}
\label{Prho0test}
\end{figure}

Numerical results for the probability distribution of $\rho_0$ for $M=1,2$ scattering channels are shown in Fig.\ \ref{Prho0test}. The agreement with the analytical distribution \eqref{PDrho0} is excellent, including the absence of any effect from the Majorana zero-mode.

\section{Thermopower distribution}
\label{thermodist}

We apply the general thermopower formulas \eqref{Presult} and \eqref{Presulteh} to a single-channel point contact, with transmission probability $T$ into the edge mode of the superconductor. There are two independent delay times $D_{1},D_{2}$ in class C, each with a twofold spin degeneracy and a twofold electron-hole degeneracy ($d_T=4$). Because of this degeneracy the class-C edge mode contains Kramers pairs of Majorana fermions. In class D the Majorana edge mode is unpaired and all delay times are nondegenerate ($d_T=1$). The point contact contributes two and the edge mode one more, so class D has a total of three independent delay times $D_{1},D_{2},D_{3}$.

Eqs.\ \eqref{Presult} and \eqref{Presulteh} can be expressed in terms of these quantities, see App.\ \ref{details}. We denote the dimensionless thermopower by $p=(\hbar/t_0){\cal S}/{\cal S}_0$ and add a subscript C,D to indicate the symmetry class. For class C we have
\begin{equation}
p_{\rm C}=\frac{(D_2/t_0-D_1/t_0)\xi\sqrt{T(1-T)}}{1-(1-T)\cos 2\beta}.\label{pCDTbeta}
\end{equation}
The independent variables $\beta,\xi$ enter via the eigenvectors of $S_0$ and $Q$, with distribution
\begin{equation}
P(\beta,\xi)=\tfrac{3}{4}(1-\xi^2)\sin 2\beta,\;\;|\xi|<1,\;\;0<\beta<\pi/2.\label{Pbetaxi}
\end{equation}
The class-D distribution $p_{\rm D}$ has a more lengthy expression, involving three delay times, see App.\ \ref{details}. These are all averages in the grand-canonical ensemble, without including effects from the charging energy of the quantum dot (which could force a transition into the canonical ensemble).\cite{Bro97b} 

\begin{figure}[tb]
\centerline{\includegraphics[width=0.9\linewidth]{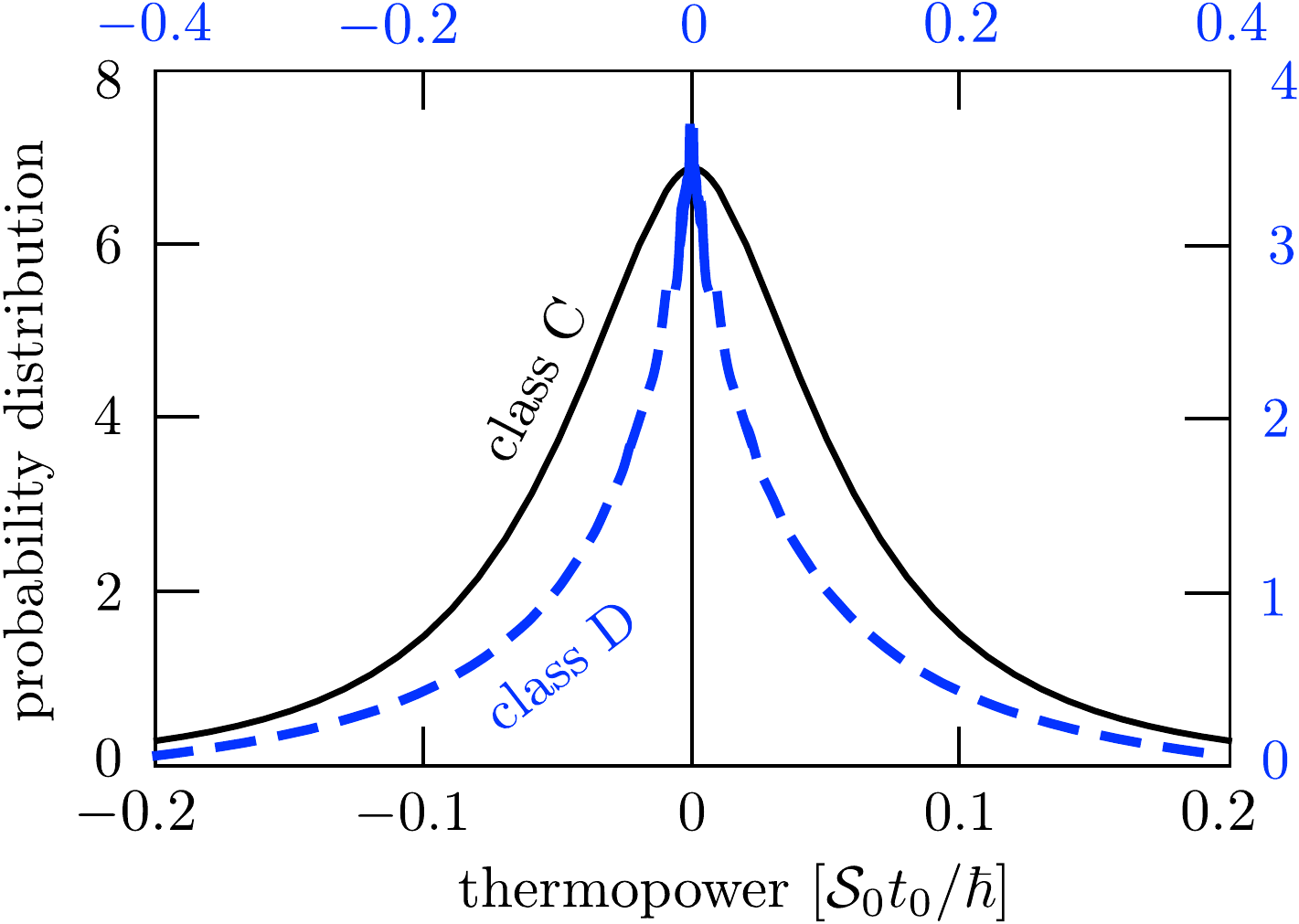}}
\caption{Probability distribution of the dimensionless thermopower $p={\cal S}\times \hbar/t_0{\cal S}_{0}$ in symmetry class C (black solid curve, bottom and left axes), and in class D (blue dashed curve, top and right axes). These are results for the quantum dot of Fig.\ \ref{fig_chiral} connecting a single-channel point contact to the unpaired Majorana edge mode of a chiral p-wave superconductor (class D), or to the paired Majorana mode of a chiral d-wave superconductor (class C).
}
\label{fig_classC}
\end{figure}

The resulting distributions, shown in Fig.\ \ref{fig_classC}, are qualitatively different, with a quadratic maximum in class C and a cusp in class D. The variance diverges in class D, while in class C
\begin{equation}
\langle p_{\rm C}^2\rangle=\frac{2}{15}(3\ln 2-2)=0.011.\label{p2C}
\end{equation}

\section{Conclusion}
\label{conclude}

\begin{figure}[tb]
\centerline{\includegraphics[width=0.8\linewidth]{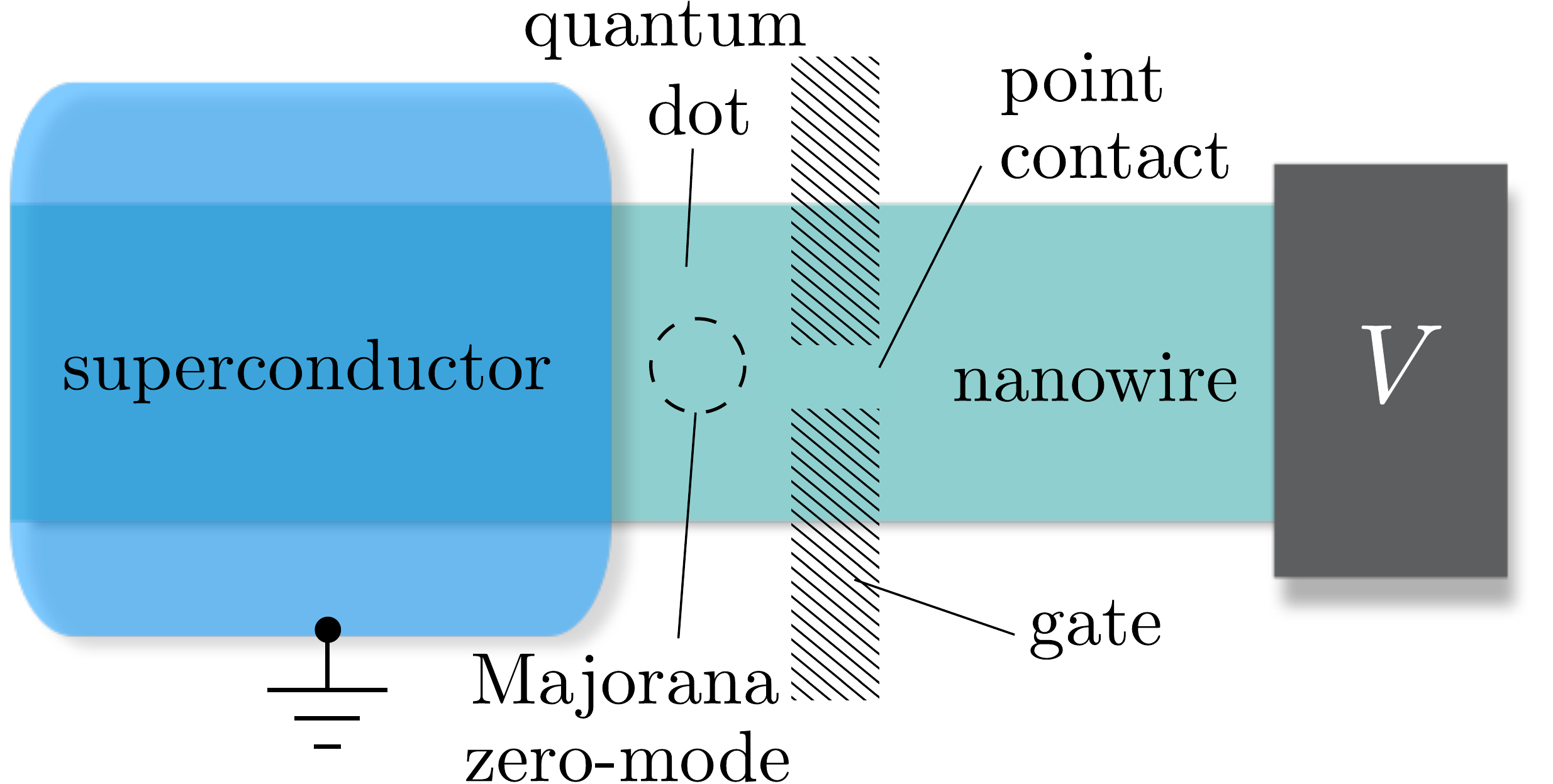}}
\caption{Geometry to detect a Majorana zero-mode by a measurement of the Andreev conductance of a ballistic point contact to a superconducting quantum dot. The probability distribution of the conductance depends on the presence or absence of the Majorana zero-mode, while the distribution of the density of states does not.
}
\label{fig_nanowire}
\end{figure}

Perhaps the most remarkable conclusion of our analysis is that the density of states of a Majorana zero-mode is not topologically protected in an open system. 

Take a superconducting quantum dot with an unpaired Majorana zero-mode and bring it into contact with a metallic contact, as in Fig.\ \ref{fig_nanowire} --- is something left of the spectral peak? The answer is ``yes'' for tunnel coupling,\cite{Bag12,Liu12,Nev13,Skv13,Sta13,Sau13,Ios13,Iva13} as it should be if the level broadening is less than the level spacing in the quantum dot. What we have found is that the answer is ``no'' for ballistic coupling, with level broadening comparable to level spacing. 

As an intuitive explanation, one might argue that this is the ultimate consequence of the fact that the two average densities of states $\rho_{+}(E)$ and $\rho_{-}(E)+\delta(E)$ of a closed quantum dot without and with a Majorana zero-mode are markedly different,\cite{Boc00,Iva02} see Eq.\ \eqref{rhopmresult}, and yet have the same integrated spectral weight of half a fermion. Still, we had not expected to find that the entire probability distribution of the Fermi-level density of states becomes identical in the topologically trivial and nontrivial system, once the quantum dot is coupled ballistically to $M\geq 1$ conducting modes.

It would be a mistake to conclude that the whole notion of a topologically nontrivial superconductor applies only to a closed system. Indeed, the Andreev conductance remains sensitive to the presence or absence of a Majorana zero-mode, even for ballistic coupling, when no trace is left in the density of states.\cite{Bee11} This can be seen most directly for the case $M=2$ of a superconducting quantum dot coupled to a normal metal by a pair of spin-resolved electron-hole modes. The Andreev conductance is then given simply by
\begin{equation}
G=\frac{e^{2}}{h}(1-{\rm Det}\,S_0),\label{GDetS0relation}
\end{equation}
and so is in one-to-one relationship with the topological quantum number ${\rm Det}\,S_0=\pm 1$. In contrast, the Fermi-level density of states has the same probability distribution \eqref{PDrho0} regardless of the sign of ${\rm Det}\,S_0$.

We have applied our results for the probability distribution of the time-delay matrix to a calculation of the thermopower induced by edge modes of a chiral p-wave or chiral d-wave superconductor.\cite{Hou13} The search for electrical edge conduction in such topological superconductors, notably ${\rm Sr}_{2}{\rm RuO}_{4}$,\cite{Mac03} has remained inconclusive,\cite{Led14} in part because of the charge-neutrality of an unpaired Majorana mode at the Fermi level.\cite{Fur01,Sto04,Ser10,Sau11} Fig.\ \ref{fig_classC} shows that both unpaired and paired Majorana edge modes can produce a nonzero thermopower --- of random sign, with a magnitude of order ${\cal S}_{0}/\delta_0=(0.3\,{\rm mV/K})\times k_{\rm B}T_0/\delta_0$. This is a small signal, but it has the attractive feature that it directly probes for the existence of propagating edge modes --- irrespective of their charge neutrality.

\acknowledgments

This research was supported by the Foundation for Fundamental Research on Matter (FOM), the Netherlands Organization for Scientific Research (NWO/OCW), the Alexander von Humboldt Foundation, and an ERC Synergy Grant.

\appendix
\section{Derivation of the delay-time distribution for the Altland-Zirnbauer ensembles}
\label{detailsdelaytime}

Repeating the steps of Refs.\ \onlinecite{Bro97} and \onlinecite{Bro99} we extend the calculation of the joint distribution $P(S_0,Q)$ from the nonsuperconducting Wigner-Dyson ensembles to the superconducting Altland-Zirnbauer ensembles. We treat the two symmetry classes C, D without time-reversal symmetry, of relevance for the main text (see Table \ref{table_CD}), and for completeness also consider the time-reversally symmetric classes CI and DIII (see Table \ref{table_CIDIII}).

\begin{table}
\centering
\begin{tabular}{ | l || c | c |}
\hline
symmetry class &  CI & DIII \\ \hline
$S$-matrix space & symplectic & orthogonal \\
& \& symmetric & \& selfdual \\ \hline
\qquad\qquad $d_T$ & $4$ &  $2$ \\ \hline
\qquad\qquad $d_E$ & $2$ &  $2$ \\ \hline
\qquad\qquad $\alpha$ & $1$&$-1$\\ \hline
\qquad\qquad $\beta$ &$2$&$2$\\ \hline
\end{tabular}
\caption{The two Altland-Zirnbauer classes with time-reversal symmetry.}
\label{table_CIDIII}
\end{table}

\subsection{Unitary invariance}
\label{invariance_proof}

Since the entire calculation relies on the unitary invariance \eqref{PSEinvariance} of the Altland-Zirnbauer circular ensembles, we demonstrate that first. Following Ref.\ \onlinecite{Bro99} we construct the ${\cal M} \times {\cal M}$ energy-dependent unitary scattering matrix $S(E)$ in terms of an ${\cal M}_0\times {\cal M}_0$ energy-independent unitary matrix $U$,
\begin{equation}
  S(E) = {\cal P} U (e^{-2 \pi i E/M_0 \delta_0} +  {\cal R} U)^{-1}{\cal P}^{\rm T}. \label{eq:SU1}
\end{equation}
The rectangular ${\cal M}\times {\cal M}_0 $ matrix ${\cal P}$ has elements ${\cal P}_{nm} =\delta_{nm}$ and ${\cal R} = 1 - {\cal P}^{\rm T}{\cal P}$. The eigenvalues $e^{i\phi_n}$ of $U$ have the same degeneracy $d_E$ as the energy eigenvalues, so there are $M_0={\cal M}_0/d_E$ distinct eigenvalues on the unit circle, arranged symmetrically around the real axis.

The ${\cal M}_0\times {\cal M}_0$ Hermitian matrix $H$ is related to $U$ via a Cayley transform,
\begin{equation}
\begin{split}
U = e^{2\pi i\epsilon/M_0\delta_0}\frac{\pi H/M_0\delta_0+i}{\pi H/M_0\delta_0-i}\\
\Leftrightarrow H=\frac{i M_0 \delta_0}{\pi}\,\frac{U+e^{2\pi i\epsilon/M_0\delta_0}}{U-e^{2\pi i\epsilon/M_0\delta_0}}.
\end{split}
\label{eq:H0def}
\end{equation}
The factor $e^{2\pi i\epsilon/M_0\delta_0}$ with $\epsilon\rightarrow 0$ is introduced to regularize the singular inverse when $U$ has an eigenvalue pinned at $+1$, as we will discuss in just a moment.

We can immediately observe that if we take a circular ensemble for $U$, with distribution function $P(U)=P(U'U)=P(UU')$, then the unitary invariance \eqref{PSEinvariance} of the distribution functional $P[S(E)]$ is manifestly true. So what we have to verify is that the construction \eqref{eq:SU1}--\eqref{eq:H0def} with $U$ in the circular ensemble is, firstly, equivalent to the Weidenm\"{u}ller formula \eqref{SHeq}, and secondly, produces a Gaussian ensemble for $H$. It is sufficient if the equivalence holds in the low-energy range $|E|\lesssim M\delta_0\ll M_0\delta_0$.

Firstly, substitution of Eq.\ \eqref{eq:H0def} into Eq.\ \eqref{eq:SU1} gives
\begin{align}
  S(E) &=
  \frac{1 + i {\cal P} \frac{M_0 \delta_0 - i \pi H \tan(\pi E^{+}/M_0
\delta_0)}{\pi H - M_0 \delta_0
        \tan(\pi E^{+}/M_0 \delta_0)} {\cal P}^{\rm T}}
       {1 - i {\cal P} \frac{M_0 \delta_0 - i \pi H \tan(\pi E^{+}/M_0
\delta_0)}{\pi H - M_0 \delta_0
        \tan(\pi E^{+}/M_0 \delta_0)} {\cal P}^{\rm T}}
  \nonumber \\ 
  &=  \frac{1 + i {\cal P} \frac{M_0 \delta_0}{\pi(H-E^{+})} {\cal P}^{\rm T}} {1 -
i {\cal P} \frac{M_0 \delta_0}{\pi(H-E^{+})} {\cal P}^{\rm T}} + {\cal O}(M/M_0),
  \label{eq:SU2}
\end{align}
with $E^{+}=E+\epsilon$. This is the Weidenm\"uller formula \eqref{SHeq}, with the ballistic coupling matrix $W = {\cal P}^{\rm T} (M_0\delta_0 /\pi^2)^{1/2}$ from Eq.\ \eqref{Wnmdef}. 

Secondly, the Cayley transform \eqref{eq:H0def} produces a Lorentzian instead of a Gaussian distribution for $H$, but in the low-energy range the two ensembles are equivalent.\cite{Bro95} One also readily checks that a uniform distribution with spacing $2\pi/M_0$ of the distinct eigenphases $\phi_n$ of $U$ produces a mean spacing $\delta_0$ of the distinct eigenvalues $E_n$ of $H$, through the relation $(\pi/M_0\delta_0)E_{n}={\rm cotan}\,(\phi_n/2)\approx (\pi-\phi_n)/2$ in the low-energy range.

The finite-$\epsilon$ regularization is irrelevant in the class C and CI circular ensembles, because there the $U$'s with an eigenvalue $+1$ are of measure zero. In the class D and DIII circular ensembles, in contrast, an eigenvalue may be pinned at unity and the regularization is essential. Let us analyze this for class D (the discussion in class DIII is similar). The matrix $U$ in class D is real orthogonal, with determinant ${\rm Det}\,U=(-1)^{n_{\rm M}}$ fixed by the parity of the number of  Majorana zero-modes [cf.\ Eq.\ \eqref{DetSonM}]. This implies that $U$ has an eigenvalue pinned at $+1$ if $M_0$ is even and $n_{\rm M}$ is odd, or if $M_0$ is odd and $n_{\rm M}$ is even. The Cayley transform \eqref{eq:H0def} then maps to an eigenvalue of $H$ at infinity. This eigenvalue does not contribute to the low-energy scattering matrix \eqref{eq:SU2}, so that it can be removed from the spectrum of $H$. Hence, whereas the dimension $M_0$ of the unitary matrix $U$ can be arbitrary, the dimension of $H$ is always even for even $n_{\rm M}$ and odd for odd $n_{\rm M}$.

\subsection{Broken time-reversal symmetry, class C and D}
\label{brokenTRS}

We now proceed with the calculation of the distribution of the time-delay matrix, first in symmetry classes C and D. Starting point is the Weidenm\"{u}ller formula \eqref{SHeq} or \eqref{eq:SU2} for the energy-dependent scattering matrix. Differentiation gives the time-delay matrix defined in Eq.\ \eqref{Qdef},
\begin{align}
&Q^{-1}=\frac{1}{2\pi\hbar}\lim_{\epsilon\rightarrow 0}[1-i\pi W^{\dagger}(H-\epsilon)^{-1}W]\nonumber\\
&\quad\times\frac{1}{W^{\dagger}(H-\epsilon)^{-2}W}[1+i\pi W^{\dagger}(H-\epsilon)^{-1}W],\label{QHrelation}
\end{align}
in terms of the Hamiltonian $H$ of the closed quantum dot and the coupling matrix $W$ to the scattering channels. The dimensionality of $H$ is $d_{E}M_0\times d_{E}M_0$ while the dimensionality of $Q$ and $S$ is $d_{T}M\times d_{T}M$ (and $W$ has dimension $d_{E}M_0\times d_{T}M$). The unitary invariance \eqref{PSEinvariance} implies $P(S_0,Q)=P(-1,Q)$, so we may restrict ourselves to the case that $H$ has a zero-eigenvalue with multiplicity $d_{T}M$ --- since then $S_0=\lim_{E\rightarrow 0}S(E)=-1$.

Restricting $H$ to its $d_{T}M$-dimensional nullspace we have, using the ballistic coupling matrix \eqref{Wnmdef},
\begin{align}
&W^{\dagger}(H-\epsilon)^{-p}W\rightarrow(M_{0}\delta_0/\pi^2)(-\epsilon)^{-p}\tilde{\Omega}^{\dagger}\tilde{\Omega},\label{Wnullspace}\\
&Q^{-1}\rightarrow(\delta_0/2\pi\hbar)\Omega^{\dagger}\Omega,\;\;\Omega=M_0^{1/2}\tilde{\Omega}.\label{QOmegarelation}
\end{align}
The matrix $\Omega$ is a $d_{T}M\times d_{T}M$ submatrix of a $d_{E}M_0\times d_{E}M_0$ unitary matrix, rescaled by a factor $\sqrt{M_0}$. In the relevant limit $M_0/ M\rightarrow\infty$ this matrix has independent Gaussian elements,
\begin{equation}
\begin{split}
P(\Omega)\propto&\exp\bigl[-\tfrac{1}{2}\beta(d_{E}/d_{T})\,{\rm Tr}'\,\Omega^{\dagger}\Omega\bigr]\\
&=\exp\bigl(-\tfrac{1}{2}\beta t_0\,{\rm Tr}'\,Q^{-1}\bigr),
\end{split}
\label{POmegadef}
\end{equation}
with $t_0=(2\pi\hbar/\delta_0)(d_E/d_T)$. The prime in the trace, and in the determinants appearing below, indicates that the $d_{T}$-fold degenerate eigenvalues are only counted once. The symmetry index $\beta$ counts the number of independent degrees of freedom of the matrix elements of $\Omega$, real in class D ($\beta=1$) and quaternion in class C ($\beta=4$). The positive-definite matrix $Q^{-1}$ of the form \eqref{QOmegarelation} is called a Wishart matrix in random-matrix theory.\cite{Forrester}

Using Eq.\ \eqref{SHeq}, an infinitesimal deviation of $S_0$ from $-1$ can be expressed as
\begin{equation}
V\Omega(S_0+1)\Omega^{\dagger}V^{\dagger}= A,\label{S0tildeH}
\end{equation}
with $A$ a $d_{T}M\times d_{T}M$ anti-Hermitian matrix, $A=-A^{\dagger}$. The matrix $A$ is a submatrix of $iH$, so its matrix elements are real in class D and quaternion in class C. The unitary matrix $V$ has been inserted so that $P(A)={\rm constant}$ near $A=0$. Since the transformation $\Omega\mapsto V\Omega$ has no effect on $P(\Omega)$ and leaves $Q$ unaffected, we may in what follows omit $V$.

The joint distribution $P(S_0,Q^{-1})$ follows from $P(\Omega)P(A)$ upon multiplication by two Jacobian determinants,
\begin{align}
&P(S_0,Q^{-1})=P(\Omega)P(A)\left|\left|\frac{\partial\Omega}{\partial Q^{-1}}\right|\right|\times\left|\left|\frac{\partial A}{\partial S_0}\right|\right|\nonumber\\
&\propto\exp(-\tfrac{1}{2}\beta t_0\,{\rm Tr}'\,Q^{-1})\left|\left|\frac{\partial \Omega^{\dagger}\Omega}{\partial\Omega}\right|\right|^{-1}\left|\left|\frac{\partial \Omega^{-1} A\Omega^{\dagger-1}}{\partial A}\right|\right|^{-1}.\label{PS0Qjacobians}
\end{align}
The Jacobians can be evaluated using textbook methods,\cite{Forrester,Mathai}
\begin{align}
&\left|\left|\frac{\partial \Omega^{\dagger}\Omega}{\partial\Omega}\right|\right|^{-1}\propto({\rm Det}'\,\Omega^{\dagger}\Omega)^{-1+\beta/2},\label{firstjacobian}\\
&\left|\left|\frac{\partial \Omega^{-1} A\Omega^{\dagger-1}}{\partial A}\right|\right|^{-1}\propto({\rm Det}'\,\Omega^{\dagger}\Omega)^{\alpha+1+(M-1)\beta/2}.\label{secondjacobian}
\end{align}
Here $\alpha+1$ equals the number of degrees of freedom of a diagonal element of $A$, while an off-diagonal element has $\beta$ degrees of freedom. So $\alpha+1=0$, $\beta=1$ for a real antisymmetric matrix $A$ (class D), while $\alpha+1=3$, $\beta=4$ for a quaternion anti-Hermitian $A$ (class C).

Collecting results, we arrive at the distribution
\begin{equation}
P(S_0,Q^{-1})\propto\exp(-\tfrac{1}{2}\beta t_0\,{\rm Tr}'\,Q^{-1})\,({\rm Det}'\,Q^{-1})^{\alpha+M\beta/2}.\label{PDfinal}
\end{equation}
The distribution \eqref{Pgamman} of the eigenvalues $\gamma_n$ of $Q^{-1}$ follows upon multiplication by one more Jacobian, from matrix elements to eigenvalues. 

\subsection{Preserved time-reversal symmetry, class CI and DIII}
\label{unbrokenTRS}

The time-reversal operator acts in a different way in class CI and DIII. In class CI the action is the transpose, so that $S=S^{\rm T}$, $H=H^{\rm T}$ are symmetric matrices. In class DIII these matrices are selfdual, $S=\sigma_{y}S^{\rm T}\sigma_{y}\equiv S^{\rm D}$, where the Pauli matrix $\sigma_y$ acts on the spin-degree of freedom. It is convenient to use a unified notation $\tilde{U}$ to denote the transpose $U^{\rm T}$ of a matrix in class CI and the dual $U^{\rm D}$ in class DIII. Unitary invariance of the circular ensemble then amounts to
\begin{equation}
P[S(E)]=P[\tilde{U}\cdot S(E)\cdot U],\label{PSEinvariance2}
\end{equation}
for energy-independent unitary matrices $U$.

Time-reversal symmetry allows to ``take the square root'' of the Fermi-level scattering matrix (Takagi factorization\cite{Hor85}),
\begin{equation}
S_0=\tilde{S}_{1/2}S_{1/2}.\label{Stildedef}
\end{equation}
In class DIII the sign of the determinant of $S_{1/2}$ is a topological quantum number,\cite{Ful11}
\begin{equation}
{\rm Det}\,S_{1/2}={\rm Pf}\,(i\sigma_y S_0)=\pm 1,\label{PfDetnumber}
\end{equation}
equal to $-1$ when the quantum dot contains a Kramers pair of Majorana zero-modes. The symmetrized time-delay matrix is defined in terms of this square root,
\begin{equation}
Q=-i\hbar \lim_{E\rightarrow 0}\tilde{S}_{1/2}^{\dagger}\frac{dS}{dE}S_{1/2}^{\dagger}.\label{QTRSdef} 
\end{equation}
The definition \eqref{Qdef} of the matrix $Q$ used in class C and D, without time-reversal symmetry, gives the same eigenvalues as definition \eqref{QTRSdef}, but would introduce a spurious correlation between $S$ and $Q$. With the definition \eqref{QTRSdef} the unitary invariance \eqref{PSEinvariance2} allows to equate $P(S_0,Q)=P(-1,Q)$, by taking $U= S_{1/2}^{\dagger}i\sigma_x$ in class CI and $U= S_{1/2}^{\dagger}\sigma_x$ in class DIII.

\begin{table}
\centering
\begin{tabular}{ | l || c | c | c | c | }
\hline
 & C & D & CI & DIII \\ \hline
$H_{nm}$& $iq_0+\bm{q}\cdot\bm{\tau}$ & $iq_0$ & $a\tau_x+b\tau_z$ & $ia\sigma_x+ib\sigma_z$\\
 ($n\neq m$) & $\beta=4$ & $\beta=1$ & $\beta=2$ & $\beta=2$ \\ \hline
$H_{nn}$ & $\bm{q}\cdot\bm{\tau}$ & $0$ & $a\tau_x+b\tau_z$ & $0$\\ 
& $\alpha+1=3$ & $\alpha+1=0$ & $\alpha+1=2$ & $\alpha+1=0$\\ \hline
\end{tabular}
\bigskip

\begin{tabular}{ | l || c | c | c |}
\hline
 & A & AI & AII \\ \hline
$H_{nm}$& $a+ib$ & $a$ & $q_0+i\bm{q}\cdot\bm{\sigma}$\\
 ($n\neq m$) & $\beta=2$ & $\beta=1$ & $\beta=4$  \\ \hline
$H_{nn}$ & $a$ & $a$ & $q_0$\\ 
& $\alpha+1=1$ & $\alpha+1=1$ & $\alpha+1=1$\\ \hline
\end{tabular}
\caption{Upper table: Representation of the Hamiltonian $H$ in the four Altland-Zirnbauer symmetry classes. All coefficients $q_n$, $a,b$ are real. The Pauli matrices $\bm{\tau}=(\tau_x,\tau_y,\tau_z)$ act on the electron-hole degree of freedom, while the $\bm{\sigma}$'s act on the spin degree of freedom. The symmetry indices $\beta$ and $\alpha+1$ from Tables \ref{table_CD} and \ref{table_CIDIII} count, respectively, the number of degrees of freedom of the off-diagonal and diagonal components of the Hermitian matrix $H$, in the Majorana basis for class D, DIII and in the electron-hole basis for class C, CI . For completeness and comparison, we show in the lower table the corresponding listing for the three Wigner-Dyson symmetry classes.}
\label{table_H}
\end{table}

Comparing to the derivation of the previous subsection, what changes is that the matrix elements of $\Omega$ and $A$ are equivalent to complex numbers $a+ib$, rather than being real or quaternion. Specifically, $\Omega$ has matrix elements of the form $a\sigma_0+ib\sigma_y$ in both class CI and DIII (to ensure that $\Omega^{\dagger}=\tilde{\Omega}$), while the matrix elements of $A$ are of the form $ia\sigma_x+ib\sigma_z$ in class CI and of the form $a\sigma_x+b\sigma_z$ in class DIII (to ensure that $A^{\dagger}=-\tilde{A}$). The Jacobian \eqref{firstjacobian} still applies, now with $\beta=2$, while the Jacobian \eqref{secondjacobian} evaluates to
\begin{equation}
\left|\left|\frac{\partial \Omega^{-1} A\Omega^{\dagger-1}}{\partial A}\right|\right|^{-1}\propto\begin{cases}
({\rm Det}'\,\Omega^{\dagger}\Omega)^{M+1}&\text{in class CI},\\
({\rm Det}'\,\Omega^{\dagger}\Omega)^{M-1}&\text{in class DIII}.\\
\end{cases}\label{thirdjacobian}
\end{equation}

Collecting results, we arrive at
\begin{equation}
P(S_0,Q^{-1})\propto\exp(-t_0\,{\rm Tr}'\,Q^{-1})\,({\rm Det}'\,Q^{-1})^{M\pm 1},\label{PCIDIIIfinal}
\end{equation}
with exponent $M+1$ in class CI and $M-1$ in class DIII. As before, the primed trace and determinant count degenerate eigenvalues only once. The distribution \eqref{Pgamman} of the eigenvalues $\gamma_n$ of $Q^{-1}$ follows with $\beta=2$ and $\alpha=\pm 1$. 

\section{Details of the calculation of the thermopower distribution}
\label{details}

\subsection{Invariant measure on the unitary, orthogonal, or symplectic groups}
\label{invariant}

For later reference, we record explicit expressions for the invariant measure $dU=P(\{\alpha_{n}\})\prod_{n}d\alpha_{n}$ (the Haar measure) in  parameterizations $U(\{\alpha_{n}\})$ of the unitary group ${\rm SU}(N)$, as well as the orthogonal or unitary symplectic subgroups ${\rm SO}(N)$, ${\rm Sp}(2N)$. (We will only need results for small $N$.) 

The invariant measure is determined by the metric tensor
\begin{equation}
g_{mn}=-{\rm Tr}\,U^{\dagger}(\partial U/\partial\alpha_m)U^{\dagger}(\partial U/\partial\alpha_n),\label{metricgdef} 
\end{equation}
via $P(\{\alpha_{n}\})\propto \sqrt{{\rm det}\,g}$. The function $P$ represents the probability distribution of the $\alpha_n$'s when the matrix $U$ is drawn randomly and uniformly from the unitary group (circular unitary ensemble, CUE), or from the orthogonal and symplectic subgroups (circular real and quaternion ensembles, CRE and CQE). 

For ${\rm SO}(2)$ we have trivially
\begin{equation}
R(\theta)=\begin{pmatrix}
\cos\theta&-\sin\theta\\
\sin\theta&\cos\theta
\end{pmatrix}\Rightarrow P(\theta)={\rm constant}.\label{PSO2}
\end{equation}
For ${\rm SU}(2)={\rm Sp}(2)$ we can choose different parameterizations:
\begin{subequations}
\label{PSU2}
\begin{align}
U&=\exp\bigl[i\beta(\tau_z\cos\theta+\tau_x\sin\theta\cos\phi+\tau_y\sin\theta\sin\phi)\bigr]\nonumber\\
&\Rightarrow P(\beta,\theta,\phi)\propto\sin^2\beta\sin\theta,\label{PSU21}\\
U&=e^{i\alpha\tau_z}\exp\bigl[i\beta(\tau_x\cos\phi+\tau_y\sin\phi)\bigr]\nonumber\\
&\Rightarrow P(\alpha,\beta,\phi)\propto \sin 2\beta,\label{PSU22}\\
U&=e^{i\alpha\tau_z}R(\theta)e^{i\alpha'\tau_z}\Rightarrow P(\alpha,\alpha',\theta)\propto\sin 2\theta.\label{PSU23}
\end{align}
\end{subequations}

For the group of $3\times 3$ orthogonal matrices we will use the Euler angle parameterization
\begin{align}
O_{\pm}&=\begin{pmatrix}
R(\alpha)&0\\
0&1
\end{pmatrix}\begin{pmatrix}
\pm 1&0\\
0&R(\theta)
\end{pmatrix}\begin{pmatrix}
R(\alpha')&0\\
0&1
\end{pmatrix}\nonumber\\
&\Rightarrow P(\alpha,\alpha',\theta)\propto\sin\theta.\label{PSO3}
\end{align}
The $\pm$ sign distinguishes the sign of the determinant ${\rm Det}\,O_{\pm}=\pm 1$, with ${\rm SO}(3)$ corresponding to $O_+$.

Finally, for ${\rm Sp}(4)$ we use the polar decomposition
\begin{align}
&U=\begin{pmatrix}
U_1&0\\
0&U_2
\end{pmatrix}\begin{pmatrix}
\tau_{0}\cos\theta&-\tau_{0}\sin\theta\\
\tau_{0}\sin\theta&\tau_{0}\cos\theta
\end{pmatrix}\begin{pmatrix}
\tau_0&0\\
0&U_3
\end{pmatrix}\nonumber\\
&\Rightarrow P(\theta)=\sin^3 2\theta.\label{UpolarU1U2U3U4}
\end{align}
The matrices $U_p$ are independently and uniformly distributed in ${\rm SU}(2)$, see Eq.\ \eqref{PSU2}. There are only three independent $U_p$'s, with 3 free parameters each, because one of the four blocks can be absorbed in the three others, so we have set it to the unit $\tau_0$ without loss of generality. (One can check that the total number $N(2N+1)\mapsto 10$ of free parameters of ${\rm Sp}(2N)$ agrees: $3+3+3$ from the $U_p$'s plus $\theta$ makes 10.) 

\subsection{Elimination of eigenvector components}
\label{eliminate}

The thermopower expressions \eqref{Presult} and \eqref{Presulteh} depend on the transmission eigenvalues $T_n$ and delay times $D_n$, but in addition there is a dependence on eigenvectors. Many of the eigenvector degrees of freedom can be eliminated by using the invariance of the distribution of the time-delay matrix under the unitary transformation $Q\mapsto U^{\dagger}QU$, following from Eq.\ \eqref{PSEinvariance}.

\subsubsection{Class C}
\label{eliminateC}

In class C we proceed as follows. The $4\times 4$ unitary symplectic scattering matrix $S_0$ has the polar decomposition \eqref{UpolarU1U2U3U4}, which we write in the form
\begin{align}
&S_0=\begin{pmatrix}
U_1&0\\
0&U_2
\end{pmatrix}\begin{pmatrix}
\tau_0\sqrt{1-T}&-\tau_0\sqrt{T}\\
\tau_0\sqrt{T}&\tau_0\sqrt{1-T}
\end{pmatrix}\begin{pmatrix}
\tau_{0}&0\\
0&U_3
\end{pmatrix},\label{S0Cparameters}\\
&U_n=e^{i\alpha_n\tau_z}\exp\bigl[i\beta_n(\tau_x\cos\phi_n+\tau_y\sin\phi_n)\bigr].\label{Ralphabetaphi}
\end{align}
We ignore the spin degree of freedom, which plays no role in the calculation. The remaining two-fold degeneracy of the transmission eigenvalue $T$ comes from the electron-hole degree of freedom.

The time-delay matrix is Hermitian with quaternion elements,
\begin{equation}
Q=\begin{pmatrix}
a\tau_0&q\\
q^{\dagger}&b\tau_0
\end{pmatrix},\;\;q=q_0\tau_0+iq_1\tau_x+iq_2\tau_y+iq_3\tau_z.\label{Qqparameter}
\end{equation}
With some trial and error, we found the unitary symplectic transformation
\begin{align}
&Q\mapsto U^{\dagger}QU,\;\;U=(U_0)^{2}\begin{pmatrix}
\tau_0&0\\
0& e^{-i\alpha_3\tau_z}U^{\dagger}_3 e^{i\alpha_3\tau_z}\end{pmatrix},\\
&U_0=\begin{pmatrix}
U_1&0\\
0&\tau_0
\end{pmatrix}\begin{pmatrix}
\tau_0\sqrt{1-T}&-\tau_0\sqrt{T}\\
\tau_0\sqrt{T}&\tau_0\sqrt{1-T}
\end{pmatrix},
\end{align}
that eliminates most of the eigenvector components from the class-C thermopower expression \eqref{Presulteh}. We are left with
\begin{equation}
{\cal S}/{\cal S}_{0}=-\hbar^{-1}\frac{2q_3\sqrt{T(1-T)}}{1-(1-T)\cos 2\beta_1}.\label{Presultehsimpler}
\end{equation}
The probability distribution of the eigenvector parameter $\beta_1$ follows from Eq.\ \eqref{PSU22},
\begin{equation}
P(\beta_1)=\sin 2\beta_1,\;\;0<\beta_1<\pi/2.\label{Pbeta1C}
\end{equation}

\subsubsection{Class D}
\label{eliminateD}

The algebra is simpler in class D, where the matrix elements are real rather than quaternion. We use the Euler angle parameterization \eqref{PSO3} of the $3\times 3$ orthogonal matrix $S_0$ with determinant ${\rm Det}\,S_0=\pm 1$. Substitution of the orthogonal transformation
\begin{equation}
Q\mapsto \begin{pmatrix}
R(-\alpha')&0\\
0&1
\end{pmatrix}Q\begin{pmatrix}
R(\alpha')&0\\
0&1
\end{pmatrix}\label{QtransformD}
\end{equation}
into the class-D thermopower expression \eqref{Presult} leads directly to
\begin{subequations}
\label{PclassDresults}
\begin{align}
&\frac{{\cal S}}{{\cal S}_{0}}=\frac{Q_{13}}{\hbar}\times\begin{cases}
-{\rm cotan}\,(\theta/2)&{\rm if}\;\;{\rm Det}\,S_0=+1,\\
\tan(\theta/2)&{\rm if}\;\;{\rm Det}\,S_0=-1,
\end{cases}\label{Presultsimpler}\\
&P(\theta)=\tfrac{1}{2}\sin\theta,\;\;0<\theta<\pi.\label{PthetaclassD}
\end{align}
\end{subequations}
The transmission eigenvalue is $T=\sin^2\theta$. Since $P(\theta)=P(\pi-\theta)$ the probability distribution of the thermopower does not depend on the sign of ${\rm Det}\,S_0$.

\subsection{Marginal distribution of an element of the time-delay matrix}
\label{Qelementdistr}

The two expressions \eqref{Presultehsimpler} and \eqref{Presultsimpler} for the thermopower contain a single off-diagonal element of the time-delay matrix $Q$. We can calculate its marginal distribution, using the eigenvalue distribution of Sec.\ \ref{circular} and the fact that the eigenvectors of $Q$ are uniformly distributed with the invariant measure of the symplectic group (class C) or the orthogonal group (class D).

\subsubsection{Class C}
\label{QelementC}

In class C the $4\times 4$ time-delay matrix $Q$ is diagonalized by a unitary symplectic matrix $U$,
\begin{align}
Q=U\begin{pmatrix}
D_1\tau_0&0\\
0&D_2\tau_0
\end{pmatrix}U^{\dagger}.\label{QUDC}
\end{align}
Each of the eigenvalues $D_1$ and $D_2$ of $Q$ has a two-fold degeneracy from the electron-hole degree of freedom. (As before, we can ignore the spin degree of freedom.) The matrix $U$ has the polar decomposition \eqref{UpolarU1U2U3U4}.

The quaternion $Q_{12}$ is given in this parameterization by
\begin{equation}
Q_{12}=\tfrac{1}{2}(D_1-D_2)(\sin 2\theta)U_1^{\vphantom{\dagger}} U_2^{\dagger},\label{Q12result}
\end{equation}
and since $q_3$ from Eq.\ \eqref{Qqparameter} equals $-\tfrac{1}{2}i\,{\rm Tr}\,\tau_{z}Q_{12}$, we have
\begin{equation}
q_3=\tfrac{1}{4}(D_1-D_2)(\sin 2\theta)\,{\rm Tr}\,U_0.\label{q3expression}
\end{equation}
The matrix $U_0=-i\tau_z U_1^{\vphantom{\dagger}} U_2^{\dagger}$ is uniformly distributed in ${\rm SU}(2)$. Using the invariant measures \eqref{PSU21} and \eqref{UpolarU1U2U3U4} we arrive at
\begin{equation}
\begin{split}
&q_3=\tfrac{1}{2}(D_1-D_2)\cos\beta\sin 2\theta,\\
&P(\beta,\theta)=(6/\pi)\sin^{2}\beta\sin^{3}2\theta,\;\;0<\beta,\theta<\pi/2.
\end{split}\label{Pbetatheta}
\end{equation}
The two angular variables $\beta,\theta$ can be combined into a single variable $\xi$:
\begin{equation}
\begin{split}
&q_3=\tfrac{1}{2}(D_1-D_2)\xi,\\
&P(\xi)=\tfrac{3}{4}(1-\xi^2),\;\;-1<\xi<1.
\end{split}\label{Pq3xi}
\end{equation} 
The marginal distribution of $q_3$ then follows upon integration. 

Collecting results, we have the following probability distributions for the variables appearing in the class-C thermopower:
\begin{align}
&{\cal S}/{\cal S}_{0}=t_0\hbar^{-1}\frac{(D_2-D_1)\xi\sqrt{T(1-T)}}{1-(1-T)\cos 2\beta},\\
&P(\beta)=\sin 2\beta,\;\;0<\beta<\pi/2,\\
&P(\xi)=\tfrac{3}{4}(1-\xi^2),\;\;-1<\xi<1,\\
&P(T)=6\,T(1-T),\;\;0<T<1,\\
&P(D_1,D_2)=\frac{32}{42525}(D_1-D_2)^{4}(D_1 D_2)^{-12}\nonumber\\
&\qquad\times\exp[-2/D_1-2/D_2],\;\;D_1,D_2>0,
\end{align}
where for notational convenience we measure the delay times in units of $t_0$.

\subsubsection{Class D}
\label{QelementD}

The $3\times 3$ time-delay matrix in class D is diagonalized by $Q=O_{+}\,{\rm diag}\,(D_1,D_2,D_3)O_{+}^{\rm T}$, with $O_{+}\in{\rm SO}(3)$ parameterized as in Eq.\ \eqref{PSO3}. In terms of these parameters, the matrix element $Q_{13}$ is given by
\begin{align}
&Q_{13}=X\cos\alpha+Y\sin\alpha,\nonumber\\
&X=\tfrac{1}{2}(D_1-D_2)\sin \theta' \sin 2\alpha',\label{Q13classDparam}\\
&Y=\tfrac{1}{2}\bigl[(D_3-D_2)\cos^{2}\alpha'+(D_3-D_{1})\sin^{2}\alpha'\bigr]\sin 2\theta',\nonumber\\
&P(\alpha,\alpha',\theta')=(8\pi^2)^{-1}\sin\theta',\;\;0<\alpha,\alpha'<2\pi,\;\;0<\theta<\pi.\nonumber
\end{align}
The thermopower distribution follows upon integration, using Eqs.\ \eqref{Pgamman}, \eqref{PclassDresults},  and \eqref{Q13classDparam}.


\begin{thebibliography}{99}
\bibitem{note1} The transformation \eqref{SMajoranatransf} from electron-hole basis to Majorana basis assumes that there is an even number of modes at each contact. This number is odd if the superconductor has an unpaired Majorana mode, in which case we have to work in the Majorana basis from the very beginning.
\bibitem{note4} We recall the definition of a quaternion, $q=q_0\tau_0+iq_1\tau_x+iq_2\tau_y+iq_3\tau_z$, with real coefficients $q_n$. A symplectic matrix $U$ is unitary, $UU^{\dagger}=1$, and satisfies $U^{\ast}=\tau_y U\tau_y$. Since $q^{\ast}=\tau_y q\tau_y$, a symplectic matrix is a unitary matrix with quaternion elements (just like an orthogonal matrix is a unitary matrix with real elements).
\bibitem{note2} Ref.\ \onlinecite{Bro97} uses a modified definition of the time-delay matrix, with a symmetrized energy derivative, to ensure the independence of $S_0$ and $Q$ also in the presence of time-reversal symmetry. This modification is not needed for the class C and D ensembles considered in the main text, so we can stay with the usual unsymmetrized definition \eqref{Qdef}. The more general case is considered in App.\ \ref{unbrokenTRS}.
\bibitem{note8} To understand why the degeneracies $d_E$ and $d_T$ of energy and transmission eigenvalues may differ in the presence of particle-hole symmetry, we recall that Kramers degeneracy applies to Hermitian operators that commute with an anti-unitary operator squaring to $-1$. The Hamiltonian $H$ anti-commutes with the particle-hole conjugation operator ${\cal C}$, so Kramers theorem does not apply. In contrast, the transmission matrix product $tt^{\dagger}$ commutes with ${\cal C}$, so when ${\cal C}^2=-1$ its eigenvalues have a Kramers degeneracy.
\end{thebibliography}
\end{document}